\documentstyle[aps,preprint]{revtex}
\newcommand{\be}{\begin{equation}}
\newcommand{\ee}{\end{equation}}
\newcommand{\bea}{\begin{eqnarray}}
\newcommand{\eea}{\end{eqnarray}}
\begin{document}
\baselineskip 20pt
\bibliographystyle{prsty}
\title{
Holes and chaotic pulses of traveling waves coupled to a long-wave mode
}
\author{Henar Herrero}
\address{Departamento de Matem\'aticas,
  Facultad de Ciencias Qu\'{\i}micas, Universidad de Castilla-La Mancha, 
13071 Ciudad Real, Spain}
\author{Hermann Riecke}
\address{Department of Engineering Sciences and Applied Mathematics,
Northwestern University, Evanston, IL 60208, USA }
\maketitle
\begin{abstract}
Localized traveling-wave pulses and holes, 
i.e. localized regions of vanishing wave amplitude, are investigated in a
 real Ginzburg-Landau equation coupled to a long-wave mode. 
In certain parameter regimes the pulses 
exhibit a Hopf bifurcation which leads
to a breathing motion. Subsequently the oscillations undergo 
period-doubling bifurcations and become chaotic. 
\end{abstract}
PACS: 47.20.ky, 47.27.te, 05.45.+b, 85.30.De.\par

\noindent
Keywords: chaos, localized patterns.
\bigskip
\newpage
Over the past few years localized structures have been investigated in a number of
pattern-forming systems, as varied as semiconductor devices \cite{Ni92,ScDr90}, 
electric gas discharges \cite{Am93}, binary fluid convection 
\cite{KoBe88,NiAh90,Ko94}, convection in narrow channels \cite{HeVi92} etc. 
Among them are structures consisting
of domains of different wavenumbers \cite{HeVi92,BrDe89,GrRi96}. In 
particular, the
localized traveling-wave patterns (`pulses') observed  
in binary-fluid convection have found considerable interest 
\cite{KoBe88,NiAh90,Ko94,ThFa88,MaNe90,HaPo91,Ri92,HeRi95,RiRa95}. 
The experimental results \cite{KoBe88,NiAh90,Ko94} 
lead to a number of theoretical analyses aimed at an understanding of the mechanism
of localization. Within the framework of a single complex Ginzburg-Landau equation
dispersion
was identified as an important ingredient for localization \cite{ThFa88,MaNe90,HaPo91}. 
It was, however, recognized that this equation does not suffice to explain
 various qualitative features of the pulses. 
The inclusion of a coupling of the traveling waves to an additional long-wave mode 
lead to a
detailed understanding of a number of qualitative features of the pulses 
\cite{Ri92,HeRi95,RiRa95}.
Equations of similar type as the resulting extended Ginzburg-Landau equations have been 
investigated in the context of two-layer Poiseuille flow \cite{ReRe93} and should apply
more generally to the interaction of traveling waves with a long-wave mode, e.g. 
in capillary jets with thermocapillarity \cite{XuDa85}.

In the present paper we investigate the extended Ginzburg-Landau equations further and
focus on two types of localized structures: periodically and aperiodically
 oscillating (`bright') pulses, and `holes' or `dark pulses', i.e. stable localized domains
of  the basic state within the traveling-wave state. In a previous paper the pulses were
described within the framework of two interacting fronts each connecting the stable basic state
with the coexisting stable traveling-wave state \cite{HeRi95}. A simple mechanism was identified
which leads to an interaction between the fronts. This interaction depends strongly on the 
direction of propagation of the pulse. 
In the context of binary-mixture convection it implied that these
pulses are only stable if they propagate opposite to their (linear) group velocity.
Simple arguments suggest that the localization mechanism should not
only stabilize pulses but also holes. We confirm this numerically in the present paper.
The asymptotic analysis used in \cite{HeRi95} only applies to {\it steadily}
 propagating pulses. Here we study numerically the regime 
outside the validity of that analysis and
find pulses which propagate unsteadily, with periodically and aperiodically varying velocities.
The phenomenon is quite similar to layer oscillations found in reaction-diffusion models
\cite{NiMi89,SuOh95}.

The extended Ginzburg-Landau equations introduced in Ref.\cite{Ri92,Ri96} in the context of
 binary-mixture convection are given to cubic order by
\bea
\partial_tA+(s+s_2C)\partial_xA&=&d\partial_x^2A+(a
+fC+f_2C^2+f_3\partial_xC)A+cA|A|^2+...,\label{e:caa1}\\
\partial_tC&=&\delta \partial_x^2C-\alpha C+h^{(2)}\partial_x|A|^2+
(h^{(1)}+h^{(3)}C)|A|^2+ \nonumber \\
& & ih^{(4)}(A^*\partial_xA-A\partial_xA^*)+....\label{e:cac1}
\eea
Here the amplitude $A$ denotes the traveling-wave amplitude and corresponds  to
 that appearing in the conventional Ginzburg-Landau equation, and $C$ characterizes
 a non-oscillatory, long-wave mode. In binary-mixture convection the long-wave mode corresponds
to a large-scale concentration field. Due to the wave character of the amplitude $A$ the
long-wave mode cannot only be generated by $A$ but also advected. For that reason - 
and because of the group velocity - equations (\ref{e:caa1},\ref{e:cac1})
are not of reaction-diffusion type. 
 
As in the asymptotic analysis in \cite{HeRi95} we neglect dispersion and assume all 
the coefficients to be real in order to focus on the effect of the long-wave mode. 
To allow an analytical treatment of the interaction of fronts a 
limit of weak diffusion was considered in \cite{HeRi95} which lead to
\bea
\partial_tA+ \eta^2 s_2 \partial_xA&=&\eta^2 d_2 \partial_x^2A  
-A+ cA^3-A^5+CA,\label{e:caa}\\
\partial_tC&=& \eta^4 \delta_4 \partial_x^2C-\eta^2 \alpha_2 C+ 
\eta^3 h^{(2)}_3 \partial_xA^2 + \eta^2 h^{(3)}_2 C A^2,
\label{e:cac}
\eea
with $\eta \ll 1$. The terms involving $h^{(1)}$ and $h^{(4)}$ were omitted since these 
coefficients turned out to vanish in the case considered in \cite{HeRi95}.
The equations were rescaled such that
the bifurcation parameter - which is related to the Rayleigh number in convection -
is the coefficient of the cubic term $c$.  The limit $\eta \ll 1$ allowed
 the derivation of coupled evolution equations for the velocities $v_{l,t}$ 
of the leading and  trailing
front, respectively, which together make up a pulse,
\bea
v_l &=& s_2 - \frac{\gamma}{|v_l|} + sgn(v_l) \rho , \label{e:vl}\\
v_t &=& s_2-\frac{\gamma}{|v_t|} + 
2\gamma \frac{e^{-\hat{\alpha} L/|v_l|}}{|v_l|} - sgn(v_t) \rho,\label{e:vt}\\ 
\partial_T L &=& \gamma \left(\frac{1}{v_t}-\frac{1}{v_l}\right)
-  2\gamma \frac{e^{-\hat{\alpha} L/|v_l|}}{v_l} +2 \rho. \label{e:Lvlvt}
\eea   
In (\ref{e:vl},\ref{e:vt},\ref{e:Lvlvt}) the length of the pulse is denoted by $L$ and 
\be
\gamma=\sqrt{3}h^{(2)}_3, \ \ \ \rho =\frac{\sqrt{3}}{\eta} (c-4/\sqrt{3}), \ \ \ T=\eta^2 t,
\ \ \ \hat{\alpha}=\alpha_2-h^{(3)}_2A_0^2.
\ee
Since in (\ref{e:caa},\ref{e:cac}) 
the diffusive term was considered much smaller than the advection term 
the interaction is one-sided, i.e. while the leading front is independent of 
the trailing front the latter feels the effect of the leading front. Therefore the peak in the
long-wave mode generated by the trailing front is reduced as compared to that of the
leading front. For $h^{(2)}>0$ this implies a repulsive interaction 
of the fronts and allows
the pulse to be stable if the pulse velocity is opposite to the group velocity \cite{HeRi95}. 
This argument suggests that backward drifting holes should be stable as well 
(for $h^{(2)}>0$). 
It should be noted that for (\ref{e:vl},\ref{e:vt},\ref{e:Lvlvt}) to be valid,
the front velocities cannot be too small. In particular, they cannot
change sign. 

\begin{figure}[p] 
\begin{picture}(420,500)(0,0)
\put(-50,0) {\includegraphics{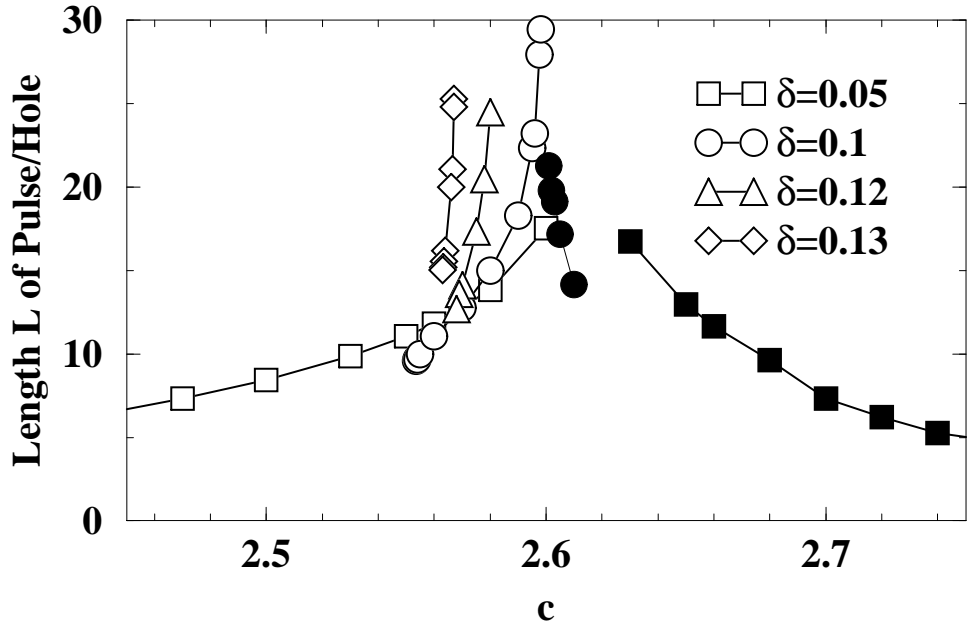}}
\end{picture}
\caption{Length of pulses (open symbols) and of holes 
(solid symbols) for different values of the diffusion coefficient of the 
long-wave mode. Open and solid squares correspond to $\delta =0.05$; open and solid
circles to $\delta =0.1$. The remaining parameters are $s=0.3$, $d=0.05$, $\alpha =0.01$,
$h^{(2)}=0.03$. 
\protect{\label{fig:Lar}}
}
\end{figure}

To investigate the stability of holes, i.e. of localized domains with vanishing amplitude $A$,
we solve (\ref{e:caa},\ref{e:cac}) for $\eta=1$ numerically using a Crank-Nicholson scheme.
We use a system with periodic boundary
conditions and a length $L=50$. The steep gradients in $C$ require 
a small grid spacing ($dx=0.025$).  The very slow dynamics allows 
large time steps ($dt=1$). As expected from the asymptotic analysis, holes are indeed
stable in a range of parameters. Their lengths are shown in fig. \ref{fig:Lar} 
(solid symbols) along with those of pulses
(open symbols). The behavior of the holes
is opposite to that of the pulses: their length increases 
with decreasing $c$ and diverges below a critical value. The region of 
existence of stable holes as well as that of pulses
decreases with increasing $\delta$, i.e. with increasing diffusion 
of the long-wave mode. Above a critical value of $\delta$ no stable holes or pulses 
are found.

\begin{figure}[p] 
\begin{picture}(420,500)(0,0)
\put(-50,0) {\includegraphics{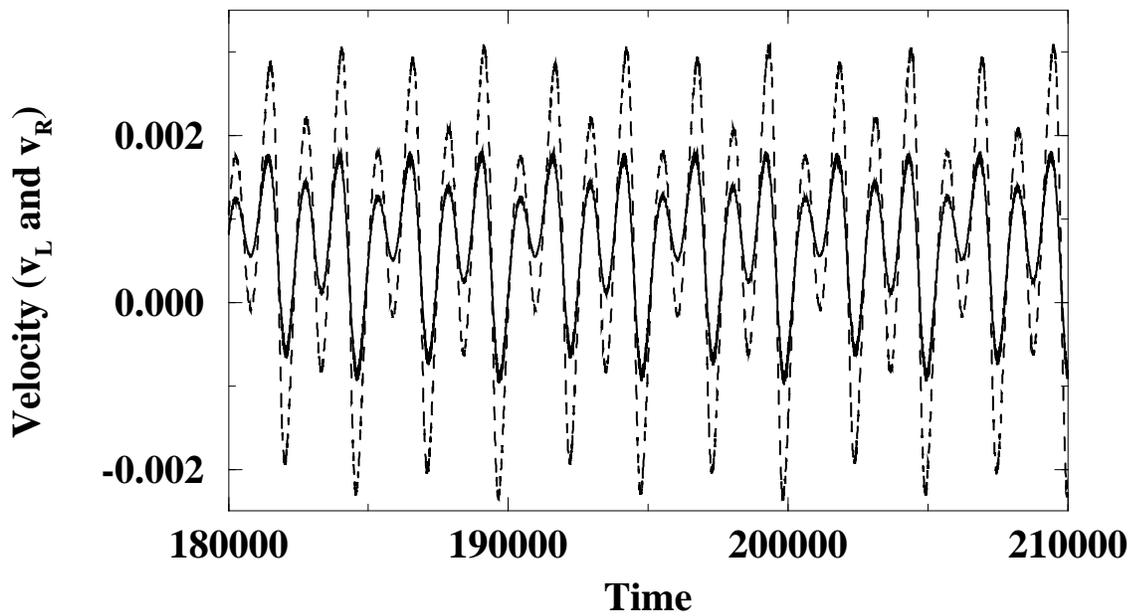}}
\end{picture}
\caption{Velocity of the left and right fronts that form the 
pulse for $c=2.561383$ and $\delta =0.13$ (period-$8$ regime). 
The remaining parameters are as in fig.\protect{\ref{fig:Lar}}.
The dashed line corresponds to the right front and the 
continuous line to the left one.
\protect{\label{fig:vlvtc2561383}}
}
\end{figure}

\begin{figure}[p] 
\begin{picture}(420,500)(0,0)
\put(-50,0) {\includegraphics{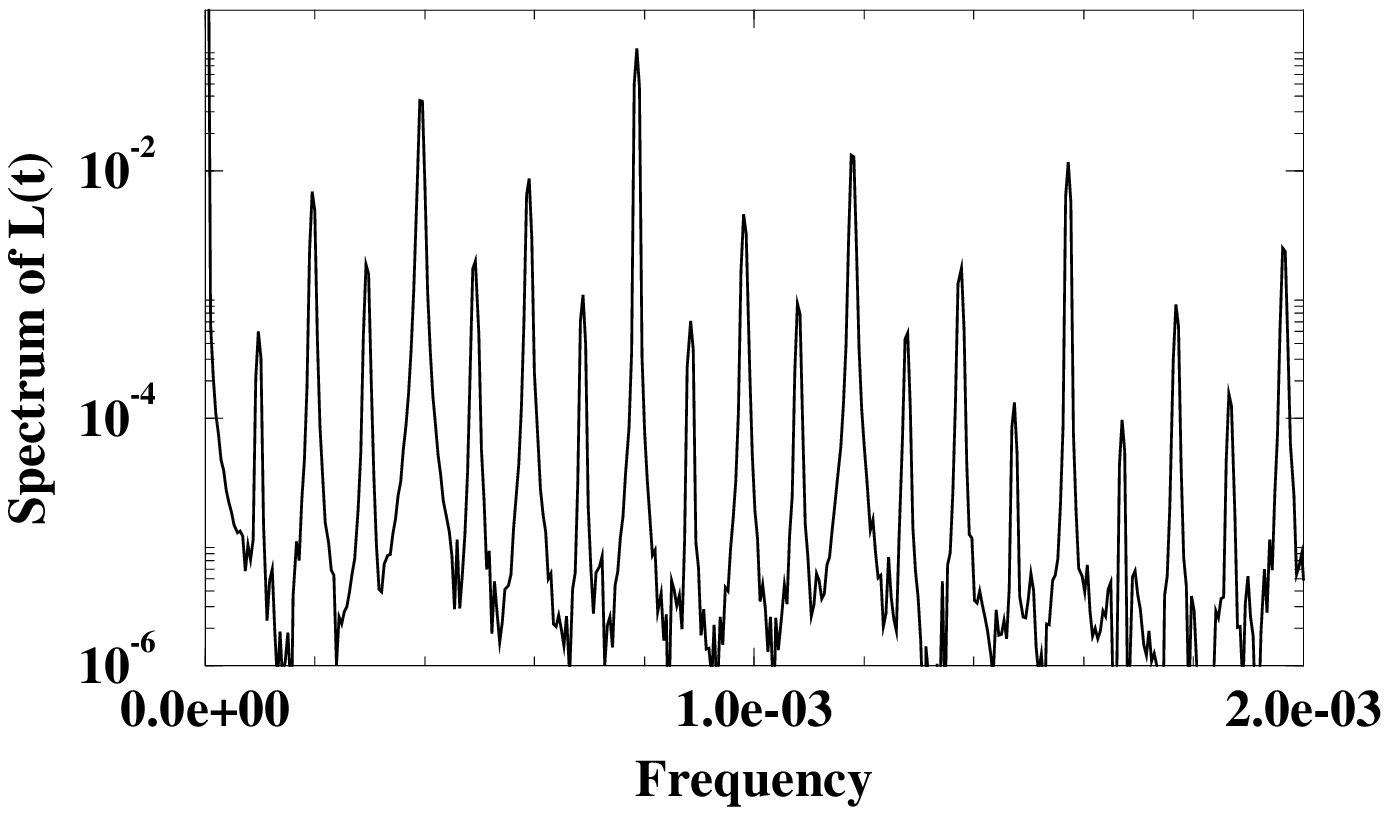}}
\end{picture}
\caption{Fourier spectrum of the length of the pulse for 
$c=2.561383$  and $\delta =0.13$ (period-$8$ regime). 
The remaining parameters are as in fig.\protect{\ref{fig:Lar}}.
\protect{\label{fig:Lfftc2561383}}
}
\end{figure}

\begin{figure}[p] 
\begin{picture}(420,500)(0,0)
\put(-50,0) {\includegraphics{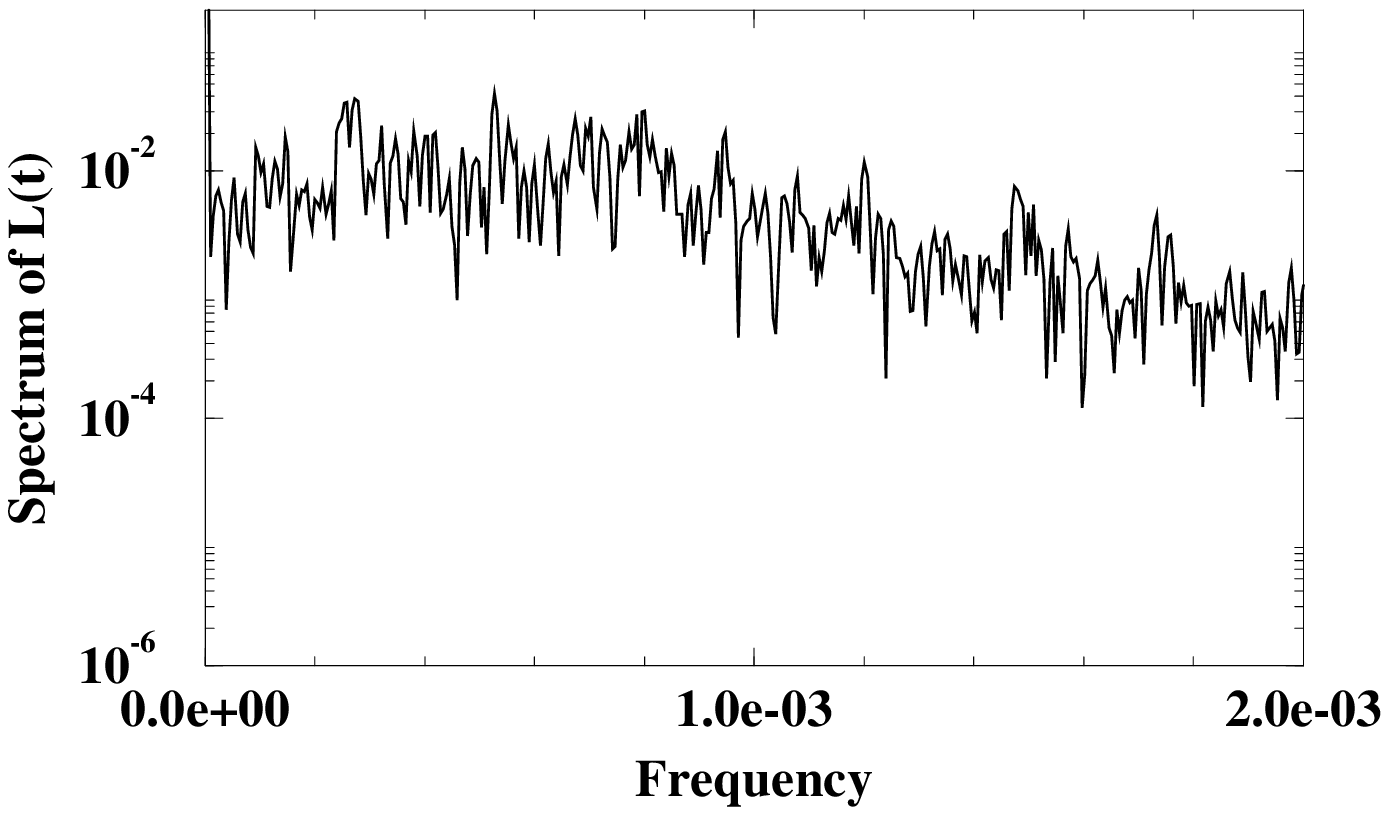}}
\end{picture}
\caption{Fourier spectrum of the length of the pulse for 
$c=2.561340$   and $\delta =0.13$ (aperiodic regime).
The remaining parameters are as in fig.\protect{\ref{fig:Lar}}.
\protect{\label{fig:Lfftc2561340}}
}
\end{figure}

\begin{figure}[p] 
\begin{picture}(420,500)(0,0)
\put(-50,0) {\includegraphics{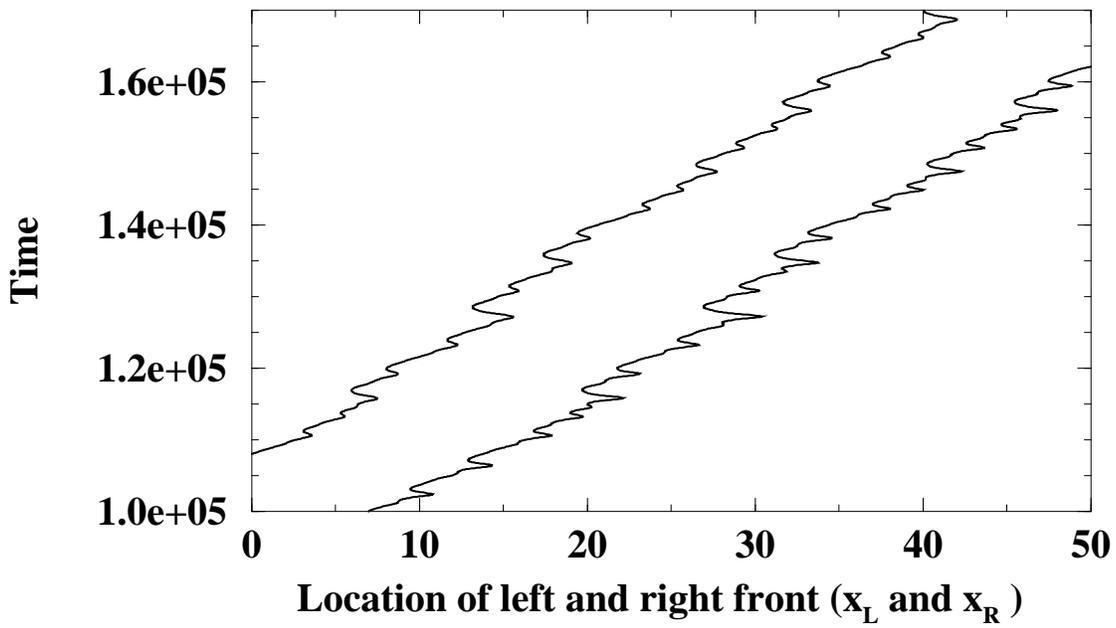}}
\end{picture}
\caption{Space-time diagram of the left and the
right front for $c=2.561340$ and $\delta =0.13$ (aperiodic regime). 
The remaining parameters are as in fig.\protect{\ref{fig:Lar}}.
\protect{\label{fig:xlxrc2561340}}
}
\end{figure}

In the weak diffusion limit of the asymptotic analysis the leading front
decouples from the trailing front and its velocity 
is independent of the
location of the trailing front. 
For finite diffusion, however,  the leading front is 
affected by the trailing front and therefore 
both fronts interact with each other.
As a consequence, more complex dynamics can be expected.
We now investigate this effect of increased diffusion on the dynamics of pulses
using the same numerical procedure.
When, for sufficiently strong diffusion, the control parameter $c$ is decreased 
a Hopf bifurcation takes place and the length and velocity of the pulse start to 
oscillate, with the fronts oscillating in antiphase. Decreasing the parameter further three period-doubling bifurcations are found to occur, leading 
to a period 8 at $c=2.561383$ (see figs.\ref{fig:vlvtc2561383} and \ref{fig:Lfftc2561383}). 
For $c=2.561380$ we observe a period 6.
For  $c=2.561340$ the dynamics become apparently chaotic 
as demonstrated in the Fourier spectrum shown in
fig.\ref{fig:Lfftc2561340}. The corresponding evolution of the 
pulse is shown in fig.\ref{fig:xlxrc2561340} which
presents a space-time diagram of the location of the two fronts making up the pulse.
Beyond this regime the
dynamics become periodic again ($c=2.561320$). Eventually, for $c<2.561310$ the pulse loses stability and grows to fill the whole system. 

\begin{figure}[p] 
\begin{picture}(420,500)(0,0)
\put(-50,0) {\includegraphics{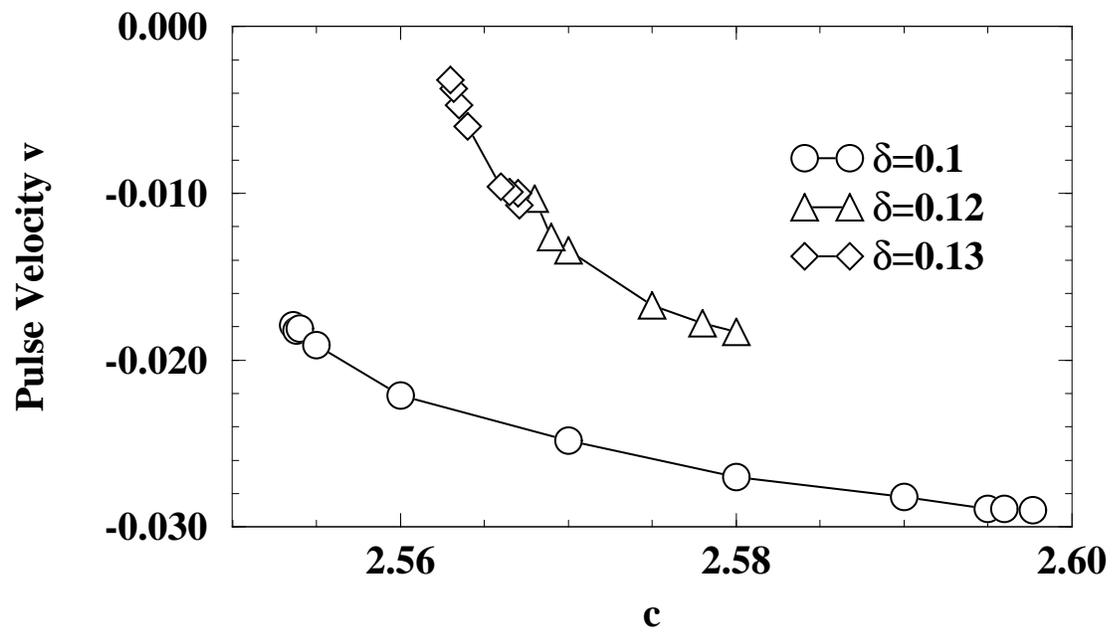}}
\end{picture}
\caption{Velocity of the pulses for different values of the 
coefficient $\delta$. 
The remaining parameters are as in fig.\protect{\ref{fig:Lar}}.
\protect{\label{fig:var}}
}
\end{figure}

For weak diffusion the (steady) pulses are unstable for positive velocity 
(in the case $h^{(2)}>0$) \cite{HeRi95} since 
the interaction between the fronts is attractive in that case.
Strikingly, for stronger diffusion the speed of the steady
pulses decreases rapidly when $c$ is decreased toward the lower end 
of the regime of stability 
(see fig. \ref{fig:var}); in 
the oscillatory pulses that arise for still smaller $c$
the instantaneous velocity can change signand the stable pulse propagates 
forward over parts of the oscillatio cycle.
In fact, in this regime even the average velocity can 
be positive (cf. fig.\ref{fig:xlxrc2561340}). 

 The oscillations shown in figs. \ref{fig:vlvtc2561383} and 
\ref{fig:xlxrc2561340} are very similar to the 
 breathing motion of localized structures  found in reaction-diffusion 
systems \cite{NiMi89,SuOh95}. They can be periodic \cite{NiMi89}  or chaotic
 \cite{SuOh95}. The systems studied in \cite{NiMi89,SuOh95} differ, however, in various aspects from that discussed here. 
Eqs.(\ref{e:caa1},\ref{e:cac1}) are not of reaction-diffusion
type since the coupling occurs $via$ the advection of the long-wave mode $C$ by the traveling-wave
mode $A$. Thus, the drift of the pulses (or holes)
is an important feature and - at least
for steady pulses - is closely related to their stability. In addition, in
\cite{SuOh95} Neumann boundary conditions were used and the distance between the two fronts making up the pulse was comparable to their distance from the boundaries. Thus, the 
boundaries may well play an important role in the dynamics. In the simulations presented
here, the system was chosen long enough to ensure that
 the (periodic) boundary conditions are not important. 
Finally, one of the three reacting 
components of the system studied in \cite{SuOh95} was assumed to diffuse very fast providing essentially a uniform background
field which gives some aspect of a global coupling. 

It should be mentioned that within the single complex Ginzburg-Landau equation 
pulses exhibiting complex dynamics have been found as well \cite{DeBr94}. In contrast to 
the pulses discussed here, these pulses
are short and stabilized by dispersion \cite{MaNe90,HaPo91,RiRa95}.
Their dynamics is related more to deformations in the
pulse shape - presumably due to phase slips - than to the size of the pulse.

In conclusion we have extended the previous analysis of  
fronts and pulses of traveling waves coupled to a long-wave mode \cite{HeRi95,HeRi94}.
Numerically, we find that not only pulses but also holes (`dark pulses') can be stabilized
by the long-wave mode. In addition, for sufficiently strong diffusion 
the mutual interaction between the leading and the trailing
front of the pulses can lead to periodic as well
as to aperiodic oscillations in the velocity and width of the pulses. No oscillatory 
holes have been found.

This work was supported by DOE through grant DE-FG02-92ER14303, by
an equipment grant from NSF (DMS-9304397) and by the DGICYT 
(Spanish Government) under grant PB93-0708.


\begin{thebibliography}{100}

\bibitem{Ni92}
F.-J. Niedernostheide, M. Arps, R. Dohmen, H. Willebrand and H.-G. Purwins, 
Phys. Status Solidi B {\bf 172}, 249 (1992).

\bibitem{ScDr90}
E. Sch\"oll and D. Drasdo, Z. Phys. B {\bf 81}, 183 (1990).

\bibitem{Am93}
E. Ammelt, D. Schweng and H.-G. Purwins, Phys. Lett. A {\bf 179}, 348 
(1993).

\bibitem{KoBe88}
P. Kolodner, D. Bensimon, and C. Surko, Phys. Rev. Lett. {\bf 60},  1723
  (1988).

\bibitem{NiAh90}
J. Niemela, G. Ahlers, and D. Cannell, Phys.~Rev.~Lett. {\bf 64},  1365
  (1990).

\bibitem{Ko94}
P. Kolodner, Phys. Rev. E {\bf 50},  2731  (1994).




\bibitem{HeVi92}
J. Hegseth, J. Vince, M. Dubois, and P. Berg\'e, Europhys. Lett. {\bf 17},  413
   (1992).

\bibitem{BrDe89}
H. Brand and R. Deissler, Phys.~Rev.~Lett. {\bf 63},  508  (1989).





\bibitem{GrRi96}
G. Granzow and H. Riecke, Phys. Rev. Lett. {\bf 77},  2451  (1996).

\bibitem{ThFa88}
O. Thual and S. Fauve, J. Phys. (Paris) {\bf 49},  1829  (1988).

\bibitem{MaNe90}
B. Malomed and A. Nepomnyashchy, Phys.~Rev. A {\bf 42},  6009  (1990).

\bibitem{HaPo91}
V. Hakim and Y. Pomeau, Eur. J. Mech. B Suppl {\bf 10},  137  (1991).

\bibitem{Ri92}
H. Riecke, Phys. Rev. Lett. {\bf 68},  301  (1992).

\bibitem{HeRi95}
H. Herrero and H. Riecke, Physica D {\bf 85},  79  (1995).

\bibitem{RiRa95}
H. Riecke and W.-J. Rappel, Phys. Rev. Lett. {\bf 75},  4035  (1995).

\bibitem{ReRe93}
M. Renardy and Y. Renardy, Phys. Fluids {\bf 5},  2738  (1993).

\bibitem{XuDa85}
J.-J. Xu and S.H. Davis. J. Fluid Mech. {\bf 161}, 1 (1985).

\bibitem{NiMi89}
Y. Nishiura and M. Mimura, SIAM J. Appl. Math. {\bf 49},  481  (1989).

\bibitem{SuOh95}
M. Suzuki, T. Ohta, M. Mimura, and H. Sakaguchi, Phys. Rev. E {\bf 52},  3645
  (1995).

\bibitem{Ri96}
H. Riecke, Physica D {\bf 92},  69  (1996).

\bibitem{DeBr94}
R. Deissler and H. Brand, Phys.~Rev.~Lett. {\bf 72},  478  (1994).

\bibitem{HeRi94}
H. Herrero and H. Riecke, Int. J. Bif. and Chaos {\bf 4},  1343  (1994).

\end{thebibliography}

\end{document}